\documentclass[onecolumn,aps,showpacs,superscriptaddress,floatfix,
               preprintnumbers,amsmath,amssymb,nofootinbib]{revtex4}

\usepackage{graphicx}
\usepackage{dcolumn}
\usepackage{bm}

\begin{document}

\title{Flipped Cryptons and UHECRs}

\author{John Ellis}
\affiliation{Theory Division, Physics Department, CERN, CH 1211 Geneva 23, 
Switzerland}
\author{V.E. Mayes}
\affiliation{George P. and Cynthia W. Mitchell Institute for Fundamental Physics, Texas A\&M 
University,\\ College Station, TX 77843, USA}
\author{D.V. Nanopoulos}
\affiliation{George P. and Cynthia W. Mitchell Institute for Fundamental Physics, Texas A\&M 
University,\\ College Station, TX 77843, USA}
\affiliation{Astroparticle Physics Group, Houston
Advanced Research Center (HARC),
Mitchell Campus,   
Woodlands, TX~77381, USA; \\
Academy of Athens,
Division of Natural Sciences, 28~Panepistimiou Avenue, Athens 10679, 
Greece}

\begin{abstract}

Cryptons are metastable bound states of fractionally-charged particles
that arise generically in the hidden sectors of models derived from
heterotic string. We study their properties and decay modes in a specific
flipped SU(5) model with long-lived four-particle spin-zero bound states
called {\it tetrons}. We show that the neutral tetrons are metastable, and
exhibit the tenth-order non-renormalizable superpotential
operators responsible for their dominant decays. By analogy with QCD, we
expect charged tetrons to be somewhat heavier, and to decay relatively
rapidly via lower-order interactions that we also exhibit. The expected
masses and lifetimes of the neutral tetrons make them good candidates for
cold dark matter (CDM), and a potential source of the ultra-high energy
cosmic rays (UHECRs) which have been observed, whereas the charged tetrons
would have decayed in the early Universe.

\end{abstract}
\pacs{04.60.-m, 11.25.Pm  ~ {\tt hep-ph/0403144} ~
ACT-02-04,CERN-PH-TH/2004-049,MIFP--04-05}
\maketitle

\section{Introduction}

Metastable particles of mass ${\cal O}(10^{12-15})$ GeV whose lifetime is
greater than the age of the Universe would be appealing cadidates for cold
dark matter, and their decays might provide the observed
ultra-high-energy cosmic rays (UHECRs)~\cite{Hayashida,HiRes}. A perfect 
candidate
for such particles is provided by `cryptons'~\cite{ELN1,ELN2,Benakli}, 
bound states
that appear in the hidden sectors of unified superstring models. It has
been pointed out that the hidden sectors of compactifications of the
heterotic string generically contain fractionally-charged
particles~\cite{Wen,Schellekens}. Since there are very stringent limits on
the abundances of fractionally-charged particles~\cite{Anthasiu}, it is
desirable to confine them, just as occurs for quarks in QCD. This is
exactly what happens to the fractionally-charged states in the flipped
SU(5) free fermionic string model~\cite{AEHN}, where this solution 
to the problem of
fractionally-charged states was first pointed out~\cite{ELN1,ELN2}, and which
remains the only example in which this solution has been worked out
in detail.

In flipped SU(5), the cryptons are bound states composed of constituents
with electric charges $\pm\frac{1}{2}$ that form $\mathbf{4}$ and
$\mathbf{\bar 4}$ representations of a hidden non-Abelian gauge group,
SO(6) $\sim$ SU(4)~\cite{AEHN}. This confines the fractionally-charged 
states into
integer-charged cryptons that may be either meson-like $\mathbf{\bar 4}
\mathbf{4}$ combinations or baryon-like states containing four
$\mathbf{4}$ or $\mathbf{\bar 4}$ states, that we term {\it tetrons}, at a
characteric mass scale $\Lambda_4 \sim
10^{11-13}$~GeV~\cite{ELN2}.  It is known that superheavy
particles $X$ with masses in the range $ 10^{11}~{\rm GeV} \lesssim M_X
\lesssim 10^{14}$~GeV might well have been produced naturally through the
interaction of the vacuum with the gravitational field during the
reheating period of the Universe following inflation in numbers sufficient
to provide superheavy dark matter~\cite{Kolb}. As was pointed out
in~\cite{Benakli}, cryptons have just the right properties to be produced
in this way, in particular because their expected masses $\sim \Lambda_4$
fall within the preferred range.

In general, tetrons may decay through $N$th-order non-renormalizable
operators in the superpotential, which would yield lifetimes that are
expected to be of the order of 
\begin{equation} 
\tau \approx \frac{\alpha_{string}^{2 - N}}{m_X} \left( \frac{M_s}{m_X} 
\right)^{2(N-3)}, 
\label{lifetime} 
\end{equation}
where $m_X$ is the tetron mass and $M_S \sim 10^{18}$~GeV is the string
scale. The $\alpha$-dependent factor reflects the expected dependence of 
high-order superpotential terms on the effective gauge coupling $g$. If 
some tetron can decay only via higher-order interactions with 
$N \geq 8$, the tetron might be much longer-lived than the age of the
Universe, in which case it might be an important form of cold dark 
matter~\cite{ELNG}.
However, no significant fraction of the astrophysical cold dark matter
could consist of charged tetrons, as these would have been detected
directly~\cite{Perl,Ambrosio,Barish}. If the neutral tetrons are close to
the experimental limit in $(m_X, \tau_X)$ space, with lifetimes in the
range $10^{15}$ years $\lesssim \tau_X \lesssim 10^{22}$
years~\cite{ELNG}, an additional possibility is that their decays might
explain the UHECRs observed by the AGASA collaboration~\cite{Hayashida},
if these turn out to exceed significantly the GZK 
cutoff~\cite{GZK,HiRes,Benakli}.

We make in this paper a detailed study of cryptons in the minimal flipped
SU(5) string model~\cite{AEHN}. A survey of non-renormalizable
superpotential terms up to tenth order enables us to investigate whether
neutral tetrons might live long enough to constitute cold dark matter, and
whether charged tetrons are likely to have had lifetimes short enough to
avoid being present in the Universe today. We also study whether the
decays of neutral tetrons could generate the UHECR. We indeed find that
charged tetrons would have decayed into neutral tetrons in the early
universe through lower-order interactions, while neutral tetrons decay
through higher-order interactions with a lifetime that makes them a
potential source for the UHECRs, as well as being attractive candidates
for cold dark matter.

\section{Field and Particle Content in the Flipped SU(5) Model}  

Already before the advent of string models, flipped $SU(5)$ attracted
interest as a grand unified theory in its own right, principally because
it did not require large and exotic Higgs representations and avoided the
straitjacket of minimal $SU(5)$ without invoking all the extra gauge
interactions required in larger groups such as $SO(10)$~\cite{Barr,DKN}.  
Interest in flipped $SU(5)$ increased in the context of string theory,
since simple string constructions could not provide the adjoint and larger
Higgs representations required by other grand unified theories. Moreover,
it was observed that flipped $SU(5)$ provided a natural `missing-partner'
mechanism for splitting the electroweak-doublet and colour-triplet fields
in its five-dimensional Higgs representations~\cite{AEHN}. We now review 
the properties of the favoured version of the flipped $SU(5)$ model 
derived from string theory, before discussing how, as an added bonus, it 
can accommodate UHECRs.

In a field-theoretic \lq flipped\rq \ $SU(5)\otimes U(1)$ model the Standard Model states occupy  $\mathbf{\bar{5}}$, $\mathbf{10}$, and $\mathbf{1}$ representations of the $\mathbf{16}$ of SO(10), with the quark and lepton assignments being \lq flipped\rq \ $u^c_L \leftrightarrow d^c_L$ and $\nu^c_L \leftrightarrow e^c_L$ relative to a conventional SU(5) GUT:
\begin{equation}
     f_{\mathbf{\bar{5}}}  =  \begin{pmatrix} u^c_1\\ u^c_2\\ u^c_3\\ e\\ \nu_e
                     \end{pmatrix}_L; \ \ \ \
     F_{\mathbf{10}}       =  \begin{pmatrix} 
                       \begin{pmatrix} u\\ d
                       \end{pmatrix}_L
                     d^c_L & \nu^c_L
                     \end{pmatrix}; \ \ \ \
     l_{\mathbf{1}} = e^c_L,
   \label{FSU5Rep}
\end{equation}
\noindent
In particular, this results in the $\mathbf{10}$ containing a neutral component with the quantum numbers of $\nu^c_L$.  Spontaneous GUT symmetry breaking can be achieved by using a $\mathbf{10}$ and $\mathbf{\bar{10}}$ of superheavy Higgs where the neutral components develop a large vacuum expectation value ({\it vev}), $ \left\langle \nu^c_H\right\rangle = 
\left\langle \bar{\nu}^c_H\right\rangle$,
\begin{equation}
H_{\mathbf{10}} = \left\{Q_H, d^c_H, \nu^c_H\right\} \ \ \ ;\ \ \  H_{\mathbf{\bar{10}}} = \left\{Q_{\bar{H}}, d^c_{\bar{H}}, \nu^c_{\bar{H}}\right\} ,
\end{equation}
while the electroweak spontaneous breaking occurs through the Higgs doublets $\mathbf{H}_2$ and $\mathbf{H}_{\bar{2}}$,
\begin{equation}
							h_{\mathbf{5}} = \left\{\mathbf{H}_2,\mathbf{H}_3\right\}\ \ \  ; \ \ \ h_{\mathbf{\bar{5}}} = \left\{\mathbf{H_{\bar{2}}}, \mathbf{H_{\bar{3}}} \right\}.
\end{equation}

The presence of a neutral component in the $\mathbf{10}$ and $\bar{\mathbf{10}}$ of Higgs fields provides a very economical doublet-triplet splitting mechanism which gives a large mass to the Higgs triplets $(\mathbf{H}_3,\mathbf{H_{\bar{3}}})$ while keeping Higgs doublets $(\mathbf{H}_2, \mathbf{H_{\bar{3}}})$ light through trilinear superpotential couplings of the form,  
\begin{eqnarray}
 FFh \rightarrow d^c_H\left\langle \nu^c_H\right\rangle H_3 \\
 \bar{F}\bar{F}\bar{h} \rightarrow \bar{d}^c_H\left\langle \bar{\nu}^c_H\right\rangle H_{\bar{3}} . 
\end{eqnarray}
Thus, in constrast to GUTs based upon other groups such as SU(5), SO(10), etc., flipped SU(5) does not require any adjoint Higgs reprentations. As a direct consequence of this, it is the only unified model that can be derived from string theory with a $k = 1$ Kac-Moody algebra~\cite{ELN1}.  
As an added bonus, this dynamic doublet-triplet splitting does not require or involve any mixing between the Higgs triplets leading to a natural suppression of dimension 5 operators that may mediate rapid proton decay. 

String-derived flipped SU(5) was created within the context of the free-fermionic formulation, which easily yelds string theories in four dimensions.  This model belongs to a class of models that correspond to compactification on 
the $Z_2\times Z_2$ orbifold at the maximally symmetric point in the Narain moduli space~\cite{Narain}.  
At the string scale, the full gauge symmetry of
the model is $ SU(5)\otimes U(1)\otimes U(1)^4 \otimes SO(10)_h \otimes
SU(4)_h $, and the spectrum contains the following massless 
fields~\cite{AEHN}.

(i) {\it Observable sector}: 

This comprises three $\mathbf{16}$ representations of SO(10), that contain
the $SU(5)\otimes U(1)$ chiral multiplets $F_i (\mathbf{10},\frac{1}{2}),\
\overline{f}_i(\overline{\mathbf 5}, \-\frac{3}{2}),\
l^c_i(\mathbf{1},\-\frac{1}{2}) (i = 1, 2, 3)$; extra matter fields
$F_4(\mathbf{10},\frac{1}{2})$, $f_4(\mathbf{5},\frac{3}{2})$,
$\bar{l}^c_4(\mathbf{1},-\frac{5}{2})$ and
$\bar{F}_5(\overline{\mathbf{10}}, -\frac{1}{2})$,
$\bar{f}_5(\bar{\mathbf{5}}, -\frac{3}{2})$, $l^c_5(\mathbf{1},
\frac{5}{2})$; and four Higgs-like fields in the $\mathbf{10}$
representation of SO(10), that $\supset SU(5)  \otimes U(1)$ representions
$h_i(\mathbf{5}, -1)$, $\bar{h}_i(\bar{\mathbf{5}},1)$, $i = 1, 2, 3, 45$.

A viable string derived flipped SU(5) model must contain the Standard Model in its light,
low-energy spectrum, whilst all other observable fields should have masses
sufficiently high to have avoided production at particle accelerators or
observation in cosmic rays. Additionally, there must be two light Higgs doublets. As we discuss below, these two objectives have been achieved in some specific variants~\cite{LoNan2,ELLN97} of the flipped SU(5)  
model, although exactly the flavor assignments of these states corresponding to those of
the standard model particle content is rather model-dependent.  However, 
a convenient choice for the flavour assignments of the fields up to mixing effects
is as follows:
\begin{eqnarray}
	\bar{f}_1: \bar{u}, \tau; \ \ \bar{f}_2: \bar{c}, e/\mu; \ \ \bar{f}_5: \bar{t}, \mu/e \\
	F_3: Q_2, \bar{s}; \ \ F_3: Q_1, \bar{d}; \ \ F_4: Q_3, \bar{b} \\
	l^c_1: \bar{\tau}; \ \ l^c_2: \bar{e}; \ \ l^c_5: \bar{u}.  
\end{eqnarray}

(ii) {\it Singlets}:  

There are ten gauge-singlet fields $\phi_{45}$, $\phi^+$, $\phi^-$,
$\phi_i (i = 1, 2, 3, 4)$, $\Phi_{12}$, $\Phi_{23}$, $\Phi_{31}$, their
ten \lq barred\rq \ counterparts, and five extra fields $\Phi_I (I = 1
\cdots 5)$. 

(iii) {\it Hidden Sector}:  

This contains 22 matter fields in the following representations of 
$SO(10)_h \otimes SU(4)_h$:
$T_i(\mathbf{10},\mathbf{1})$, $\Delta_i(\mathbf{1}, \mathbf{6}) (i = 1
\cdots 5)$; $\tilde{F}_i(\mathbf{1}, \mathbf{4})$,
$\tilde{\bar{F}}_i(\mathbf{1}, \mathbf{\bar{4}})(i = 1 \cdots 6)$. Flat
potential directions along which the anomalous combination of 
hypercharges $U(1)_A$ is cancelled
induce masses that are generally near the string scale
for some, but not all, of these states.  Depending upon the
number of $T_i$ and $\Delta_i$ states remaining massless, the SO(10)
condensate scale is $10^{14-15}$ GeV and the SU(4) condensate scale is 
$10^{11-13}$~\cite{LN1996}
GeV.  The $\tilde{F}_{3,5}$ and $\tilde{\bar{F}}_{3,5}$ states always
remain massless down to the condensate scale.  The $U(1)_i$ charges and
hypercharge assignments are shown in the Table below.

In order to preserve D and F flatness, many of the singlet fields can
develop vacuum expectation values, as can some of the hidden-sector 
fields. Many of these
flat directions have been studied in detail~\cite{CEN2000}. Typically, we
have $\left\langle \Phi_{23}, \Phi_{31}, \bar{\Phi}_{23}, \bar{\Phi}_{31},
\phi_{45}, \bar{\phi}_{45}, \phi^+, \phi^- \right\rangle \neq 0$, while it can be shown that there is no solution unless
$\left\langle \Phi_3, \Phi_{12}, \bar{\Phi}_{12} \right\rangle = 0$. The
phenomenological details of a particular model depends upon the flat direction which
is chosen.  

The superheavy Higgs $H_{\mathbf{10}}$ can in general be a linear combination of $F_1$, $F_2$, $F_3$, and $F_4$ , while $H_{\mathbf{\bar{10}}} = \bar{F}_5$.  The Higgs doublet matrix takes the following form, including terms up to $5th$ order in the superpotential:
\begin{equation}
      m_h = \begin{pmatrix} 0 & \Phi_{12} & \bar{\Phi}_{31} & T^2_5\bar{\phi}_{45} \\
                            \bar{\Phi}_{12} & 0 & \Phi_{23} & \Delta^2_4\bar{\phi}_{45} \\
                            \Phi_{31} & \bar{\Phi}_{23} & 0 & \bar{\phi}_{45} \\
                            \Delta^2_5 & T^2_4\phi_{45}& \phi_{45} & 0 
            \end{pmatrix}
\end{equation}
If only all-order contributions generated by singlet vevs are considered, $H_1$, $H_{245} = cos \ \theta H_2 - sin \ \theta H_{45}$, $\bar{H}_{12} = cos \ \bar{\theta}\bar{H}_1 - sin \ \bar{\theta}\bar{H_2}$, and $\bar{H}_{45}$ light, where $tan \ \theta = \left\langle \Phi_{23}\right\rangle/\left\langle \phi_{45}\right\rangle$ and $tan \ \bar{\theta} = \left\langle \Phi_{31}\right\rangle/\left\langle \bar{\Phi}_{23}\right\rangle$.  The $\left\langle TT\right\rangle$ in the Higgs doublet matrix give additional structure. 
With the choice $\left\langle \Phi_{12},\bar{\Phi}_{12}\right\rangle = 0$ and with the additional constraints $\left\langle \Delta^2_i \right\rangle = 0$ and $\left\langle T^2_i \right\rangle = 0$, the massless Higgs doublet eigenstates are
identified as $H_{\mathbf{2}} = H_1$ and $H_{\mathbf{\bar{2}}} = \bar{H}_{45}$.  
Similarly, the Higgs triplet mass matrix can be formed, and it is found that all of the Higgs triplets become massive~\cite{LoNan2}.

If the state $F_{\beta} \propto -\left\langle F_3 \right\rangle F_1 + \left\langle F_1 \right\rangle F_3$ is the linear combination that does not receive a vev, the flavour identification of the quarks and leptons with the specific string representations can be made:
\begin{equation}
t \ b \ \tau \ \nu_{\tau}: \ \ \  Q_4 \ d^c_4 \ u^c_5 \ L_1 \ l^c_1,
\end{equation}
\begin{equation}
c \ s \ \mu \ \nu_{\mu}: \ \ \ Q_2 \ d^c_2 \ u^c_2 \ L_2 \ l^c_2,
\end{equation}
\begin{equation}
u \ d \ e \ \nu_{e}: \ \ \ Q_{\beta} \ d^c_{\beta} \ u^c_1 \ L_5 \ l^c_5.  
\end{equation}

In addition to the above states which have been identified with those of the Standard Model, there are extra states $\bar{f}_3$ and $l^c_3$, as well as \lq exotic\rq \ states $f_4$ and $\bar{l}^c_4$ which should not appear in the light 
spectrum.  In particular, there are $5th$ order superpotential terms that contain $\bar{f}_3$ and $l^c_3$ which can generate dimesion-five operators leading to rapid proton decay.  Fortunately, there are superpotential terms~\cite{LoNan2} of the form
\begin{equation}
f_4 \sum_i \alpha_i\bar{f}_i, \ \ \ \ \  \bar{l}^c_4 \sum_i \alpha_i l^c_i
\end{equation}
which allow these states to become heavy.
  
The singlet fields also may potentially obtain masses. 
The relevant trilinear couplings involving the 
singlet fields are
\begin{equation}
\frac{1}{2}(\phi_{45}\bar{\phi}_{45}\Phi_3 + \phi^+\bar{\phi}^+\Phi_3 + 
\phi^-\bar{\phi}^- \Phi_3 + \phi_i\bar{\phi}_i\Phi_3)
+ (\phi_1\bar{\phi}_2 + \bar{\phi}_1\phi_2)\Phi_4 + 
(\Phi_{12}\Phi_{23}\Phi_{31} + \Phi_{12}\phi^+\phi^- + 
\Phi_{12}\phi_i\phi_i + h.c.),
\end{equation}
from which it is clear that having $\left\langle \Phi_3 \right\rangle \neq
0$ would give trilinear mass terms for $\phi_{45}$, $\phi^+$, $\phi^-$,
$\phi_i$ and their barred counterparts.  However, $\left\langle \Phi_3
\right\rangle = 0$ is required. Moreover, we have the result~\cite{LoNan2} 
\begin{equation}
\phi^N = 0, \ N\geq 4,
\end{equation}
Hence, we expect that most of the singlet fields will remain light.

\begin{table}
{\it {Charges and hypercharges for crypton fields in the flipped SU(5) 
model~\cite{AEHN}}}\\
{~}\\
\begin{tabular}{|r|l|c|r|r|r|c|c|}
\hline
State      & $SU(4) \otimes SO(10)$      & $U_1(1)$ & $U_2(1)$ & $U_3(1)$ 
& $U_4(1)$ \\[0.5ex]
\hline
$\Delta_1$ & $(\mathbf{6},\mathbf{1})^0$ & \ 0        & $-\frac{1}{2}$ & $\ \frac{1}{2}$ & \ 0  \\
$\Delta_2$ & $(\mathbf{6},\mathbf{1})^0$ & $-\frac{1}{2}$ &\ 0 & $\ \frac{1}{2}$ & \ 0 \\
$\Delta_3$ & $(\mathbf{6},\mathbf{1})^0$ & $-\frac{1}{2}$ & $\ -\frac{1}{2}$ & \ 0 & $\ \frac{1}{2}$ \\
$\Delta_4$ & $(\mathbf{6},\mathbf{1})^0$ & \ 0        & $-\frac{1}{2}$ & $\ \frac{1}{2}$ & \ 0  \\
$\Delta_5$ & $(\mathbf{6},\mathbf{1})^0$ & $ \frac{1}{2}$ &\ 0 & $\ -\frac{1}{2}$ & \ 0 \\
\hline
\\
\hline
$T_1$ & $(\mathbf{1},\mathbf{10})^0$ & \ 0        & $ -\frac{1}{2}$ & $\ \frac{1}{2}$ & \ 0  \\
$T_2$ & $(\mathbf{1},\mathbf{10})^0$ & $ -\frac{1}{2}$ &\ 0 & $\ \frac{1}{2}$ & \ 0 \\
$T_3$ & $(\mathbf{1},\mathbf{10})^0$ & $ -\frac{1}{2}$ & $\ -\frac{1}{2}$ & \ 0 & $\ -\frac{1}{2}$ \\
$T_4$ & $(\mathbf{1},\mathbf{10})^0$ & \ 0        & $\frac{1}{2}$ & $\ -\frac{1}{2}$ & \ 0  \\
$T_5$ & $(\mathbf{1},\mathbf{10})^0$ & $ -\frac{1}{2}$ &\ 0 & $\ \frac{1}{2}$ & \ 0 \\
\hline
\\
\hline
$\tilde{F}_1$ & $(\mathbf{4},\mathbf{1})^{+5/4}$ & $ -\frac{1}{4}$  & $ \frac{1}{4}$ & $\ - \frac{1}{4}$ & \ $ \frac{1}{2}$  \\
$\tilde{F}_2$ & $(\mathbf{4},\mathbf{1})^{+5/4}$ & $ -\frac{1}{4}$  & $ \frac{1}{4}$ & $\ - \frac{1}{4}$ & \ $ -\frac{1}{2}$  \\
$\tilde{F}_3$ & $(\mathbf{4},\mathbf{1})^{-5/4}$ & $ \frac{1}{4}$  & $ \frac{1}{4}$ & $\ - \frac{1}{4}$ & \ $ \frac{1}{2}$  \\
$\tilde{F}_4$ & $(\mathbf{4},\mathbf{1})^{+5/4}$ & $ \frac{1}{4}$  & $ -\frac{1}{4}$ & $\ - \frac{1}{4}$  & \ $ -\frac{1}{2}$  \\
$\tilde{F}_5$ & $(\mathbf{4},\mathbf{1})^{+5/4}$ & $ -\frac{1}{4}$  & $ \frac{3}{4}$ & $ \frac{1}{4}$ & \ 0  \\
$\tilde{F}_6$ & $(\mathbf{4},\mathbf{1})^{+5/4}$ & $ -\frac{1}{4}$  & $ \frac{1}{4}$ & $\ - \frac{1}{4}$ & \ $\ - \frac{1}{2}$  \\
\hline
\\
\hline
$\tilde{\bar{F}}_1$ & $(\mathbf{\bar{4}},\mathbf{1})^{-5/4}$ & $ -\frac{1}{4}$  & $ \frac{1}{4}$ & $\  \frac{1}{4}$ & \ $ \frac{1}{2}$  \\
$\tilde{\bar{F}}_2$ & $(\mathbf{\bar{4}},\mathbf{1})^{-5/4}$ & $ -\frac{1}{4}$  & $ \frac{1}{4}$ & $\  \frac{1}{4}$ & \ $ -\frac{1}{2}$  \\
$\tilde{\bar{F}}_3$ & $(\mathbf{\bar{4}},\mathbf{1})^{+5/4}$ & $ -\frac{1}{4}$  & $ -\frac{1}{4}$ & $\  \frac{1}{4}$ & \ $ -\frac{1}{2}$  \\
$\tilde{\bar{F}}_4$ & $(\mathbf{\bar{4}},\mathbf{1})^{-5/4}$ & $ -\frac{1}{4}$  & $ \frac{1}{4}$ & $\  \frac{1}{4}$  & \ $ -\frac{1}{2}$  \\
$\tilde{\bar{F}}_5$ & $(\mathbf{\bar{4}},\mathbf{1})^{-5/4}$ & $ -\frac{3}{4}$  & $ \frac{1}{4}$ & $ -\frac{1}{4}$ & \ 0  \\
$\tilde{\bar{F}}_6$ & $(\mathbf{\bar{4}},\mathbf{1})^{-5/4}$ & $ \frac{1}{4}$  & $ -\frac{1}{4}$ & $\  \frac{1}{4}$ & \ $\ - \frac{1}{2}$  \\
\hline
\end{tabular}
\end{table}

\section{Crypton Bound States}

Since the strong-interaction scale for the SU(4) factor in the hidden 
sector is expected to lie
below that for the SO(10) factor, we concentrate on the states bound by
the hidden-sector SU(4) interactions.  These include `holomorphic'
`mesons' with the contents $T_iT_j$, $\Delta_i\Delta_j$ and
$\tilde{F}_i\tilde{\bar{F}}_j$, `non-holomorphic' mesons with the contents
$T_iT_j^*$, $\Delta_i\Delta_j^*$ and $\tilde{F}_i\tilde{F}_j^*$, `baryons'
with the contents $\tilde{F}_i\tilde{F}_j\Delta_k$ and
$\tilde{\bar{F}}_i\tilde{\bar{F}}_j\Delta_k$, and quadrilinear {\it
tetrons}, with the contents of four $\tilde{F}_i$ and/or
$\tilde{\bar{F}}_i$ fields and/or their complex conjugates. We assume that
the baryons are heavier than the lightest tetrons, which are expected to
be BPS-like `holomorphic' states with the quantum numbers of
$\tilde{F}_i\tilde{F}_j\tilde{F}_k\tilde{F}_l$ and
$\tilde{\bar{F}}_i\tilde{\bar{F}}_j\tilde{\bar{F}}_k\tilde{\bar{F}}_l$,
where $i, j, k, l = 3,5$. `Non-holomorphic' tetrons with the quantum
numbers of $\tilde{F}_i\tilde{F}_j\tilde{F}_k (\bar{\tilde{F}}_l)^*$,
$\tilde{F}_i\tilde{F}_j(\bar{\tilde{F}}_k)^* (\bar{\tilde{F}}_l)^*$, etc.,  
are generally expected to be heavier, although this remains to be proved. 
We assume that, by analogy with QCD, these excited states have short 
lifetimes.

Crypton bound states occur in `cryptospin' multiplets with different
permutations of confined constituents, analogous to the flavour SU(3) and
SU(4) multiplets of bound states in QCD. We recall that the
observable-sector non-Abelian gauge interactions do not act on the
hidden-sector supermultiplets, and assume masses $\gg \Lambda_4$ for all
the U(1) gauge supermultiplets except that in the Standard Model, in which
case they also do not contribute significantly to the cryptospin mass
splittings. Two classes of diagrams  are
likely to contribute to the mass differences between `cryptospin'
partners: electromagnetic `self-energy' diagrams 
and the photon-exchange `Coulomb potential' diagrams.
We do not enter here into a discussion which of
these classes of diagrams is likely to dominate for which cryptospin
multiplets, as this is not essential for our purposes. 

We expect these diagrams to have the following orders of magnitude:
\begin{equation}
{\cal O}({\alpha \over \pi}) \Lambda_4 \times \{ {\rm (a)} \, \Sigma_i
Q_i^2, \;
{\rm (b)} \, Q_T^2 \},
\label{splittings}
\end{equation}
where the $Q_i$ are the charges of the tetron constituents, and $Q_T$ is 
the total tetron charge. It is easy to check the well-known fact that both 
of these terms make 
positive contributions to both the $\pi^+ - \pi^0$ and $p - n$ mass 
differences. The former agrees with experiment in sign and order of 
magnitude, and the difference of the 
latter from experiment is explained by the difference between the $u$ and 
$d$ quark masses, so one may have some confidence in the qualitative
estimates in (\ref{splittings}).

Each of the dependences in (\ref{splittings}) would give $m_{T^{++}} >
m_{T^{+}} > m_{T^{0}}$. We therefore expect the doubly-charged tetrons
\begin{equation}
\Psi^{--} = \tilde{F}_3\tilde{F}_3\tilde{F}_3\tilde{F}_3, \; \;
\Psi^{++} = \tilde{F}_5\tilde{F}_5\tilde{F}_5\tilde{F}_5,
\end{equation}
\begin{equation}
\bar{\Psi}^{++} = 
\tilde{\bar{F}}_3\tilde{\bar{F}}_3\tilde{\bar{F}}_3\tilde{\bar{F}}_3, \; 
\;  \bar{\Psi}^{--} = 
\tilde{\bar{F}}_5\tilde{\bar{F}}_5\tilde{\bar{F}}_5\tilde{\bar{F}}_5,
\end{equation}
to be heavier than the singly-charged states
\begin{equation}
\Psi^{+} = \tilde{F}_3\tilde{F}_5\tilde{F}_5\tilde{F}_5, \; \;
\Psi^{-} = \tilde{F}_5\tilde{F}_3\tilde{F}_3\tilde{F}_3,
\end{equation}
\begin{equation}
\bar{\Psi}^{-} = 
\tilde{\bar{F}}_3\tilde{\bar{F}}_5\tilde{\bar{F}}_5\tilde{\bar{F}}_5, \; 
\;
\bar{\Psi}^{+} = \tilde{\bar{F}}_5\tilde{\bar{F}}_3\tilde{\bar{F}}_3\tilde{\bar{F}}_3,
\end{equation}
which are in turn expected to be heavier than the neutral states
\begin{equation}
\Psi^{0} = \tilde{F}_3\tilde{F}_3\tilde{F}_5\tilde{F}_5, \; \;
\bar{\Psi}^{0} = \tilde{\bar{F}}_3\tilde{\bar{F}}_3\tilde{\bar{F}}_5\tilde{\bar{F}}_5.
\end{equation}
Just like the proton in QCD, the lowest-lying neutral tetrons can decay
only via higher-order operators in the superpotential, as we discuss
below. This may make them good candidates for cold dark matter as well as
providing via their decays a possible source of the UHECRs~\cite{Benakli}.

\section{The Decays of the lightest SU(4) Mesons}

We first discuss the decays of the lightest hidden-sector SU(4)  
bound-state mesons. In analogy with QCD chiral symmetry breaking, it is
expected that there will be an isotriplet of cryptopions that could play 
the role of
pseudo-Nambu-Goldstone bosons, with masses that are small compared to
$\Lambda_4$. Specifically, the charged SU(4) pion states
\begin{equation} 
   \pi^{\pm} = (\tilde{F}_3 \tilde{\bar{F}}_5,  \tilde{\bar{F}}_3 \tilde{F}_5).
\end{equation}  
are expected to have masses
\begin{equation}
m^2_{\pi^\pm} = \Lambda_4 \times (m_3 + m_5),
\end{equation}
where $m_{3,5}$ are the bare masses of the fractionally-charged 
constituents, which are expected to be $< \Lambda_4$, as we discussed 
above. The neutral SU(4) pion state
\begin{equation}
 \pi^0 = \frac{1}{\sqrt{2}}( \tilde{F}_3 \tilde{\bar{F}}_3 - \tilde{F}_5 
 \tilde{\bar{F}}_5).
\end{equation}
is expected to be lighter by an amount
\begin{equation}
m^2_{\pi^{\pm}} - m^2_{\pi^0} = (\frac{\alpha}{\pi})\Lambda_4^2 
ln(\Lambda_4^2/m^2_{\pi^0}).
\label{pi+pi0difference})
\end{equation}
The cryptospin-zero state
\begin{equation}
\eta^0 = \frac{1}{\sqrt{2}} (\tilde{F}_3 \tilde{\bar{F}}_3 + \tilde{F}_5\tilde{\bar{F}}_5) 
\end{equation}
is expected to be significantly heavier because of a $U_A(1)$ anomaly.  

We find that there are $N = 3$ superpotential terms of the form
\begin{eqnarray}
    \tilde{F}_3\tilde{\bar{F}}_3 \Phi_3 -  \tilde{F}_5\tilde{\bar{F}}_5\bar{\Phi}_{12}
\end{eqnarray}
that would allow the crypto-$\pi^0$ and -$\eta^0$ mesons to decay very 
rapidly.
Additionally, we expect the crypto-$\pi^0$ and -$\eta^0$ states to have 
couplings to
pairs of photon supermultiplets, analogous to those of the QCD $\pi^0$ and
$\eta^0 \to \gamma \gamma$. These couplings would be described in an 
effective supergravity lagrangian by
terms in the chiral gauge kinetic function $f$ of the form
$\alpha \pi / \Lambda_4$ and $\alpha \eta / \Lambda_4$, where $\Pi, \eta$
denote composite superfields and $\Lambda_4$ is the scale at which the
hidden-sector SU(4) interactions become strong. As in the case of the
QCD $\pi^0$ decaying to $\gamma \gamma$, these couplings would give very 
short
lifetimes for the crypto-$\pi^0$ and -$\eta^0$ states. It is also possible
that in some variant models the crypto-$\pi^0$ and -$\eta^0$ might have 
additional
decays, analogous to those of the QCD $\eta^0$, which would further
shorten their lifetimes.

In the case of the charged cryptopions, we find terms of the form
\begin{equation}
\pi^-(   F_2   F_2   F_3  \bar{h}_{45} + 
     F_3   F_4   \phi_2   f_4 +
     F_3   F_4   h_1   l^c_5  + 
     F_3  \bar{h}_{45} \bar{f}_2   l^c_2 +
     F_3  \bar{h}_{45} \bar{f}_5   l^c_5),
\label{pi-decay}
\end{equation}
and 
\begin{equation}
    \pi^+ (   F_4   \Phi_{31} \bar{f}_3  \bar{f}_5 + 
    \bar{\phi}_{45}  f_4  \bar{f}_3  \bar{l}^c_4 +
    \bar{\phi}_{45} \bar{l}^c_4 \bar{l}^c_4  l^c_3),
\end{equation}  
that would allow the $\pi^{\pm}$ states to decay fairly rapidly.

There would also be a complex spectrum of heavier `non-holomorphic' SU(4)  
bound-state mesons, analogous to the $\rho$ and heavier mesons of QCD, but
we expect them all to be very unstable, and do not discuss them further.
Likewise, we do not discuss mesons made of the higher SU(4)
representations $\Delta_i$, or $F F \Delta$ cryptobaryons, or SO(10) bound
states, as these have been studied previously in~\cite{ELN2}.

\section{The Fate of the Neutral Tetrons}
As discussed above, we expect the lightest tetrons to be the
electrically neutral states. These can decay only through higher-order 
non-renormalizable superpotential terms, for which the first 
candidates appear at eighth order:
\begin{eqnarray}
     \Psi^0 && F_4   \phi^- \bar{h}_2  \bar{f}_5, \\                  
\bar{\Psi}^0 && \phi^+ \bar{h}_{45}  f_4  \bar{l}^c_4.
\end{eqnarray}
At ninth order, we find terms involving neutral tetrons of the 
following forms:
\begin{eqnarray} 
  \bar{\Psi}^0 & ( \Phi_{31}        f_4   f_4  \bar{f}_3  \bar{f}_3  + 
     \Phi_{31}          f_4  \bar{f}_3  \bar{l}^c_4  l^c_3 + 
 \Phi_{31}          \bar{l}^c_4 \bar{l}^c_4  l^c_3  l^c_3), \\
\Psi^0  & ( F_1 \phi_1 \phi^- \bar{h}_2 \bar{f}_1  + F_2   \phi_4   
\phi^-  \bar{h}_1       \bar{f}_2 + 
F_2  \bar{\phi}_4   \phi^- \bar{h}_2  \bar{f}_2 
+  F_2 \bar{\phi}_4 \phi^- \bar{h}_2 \bar{f}_2).
\end{eqnarray}
All of these $8th$ and $9th$ order terms contain fields which are expected to have large masses, so we do not expect
that these decay modes would be kinematically accessible.  
The next terms yielding possible neutral tetron decays are of tenth order.
There are a large number of such terms, of which the following are
those containing only fields that are light in the model:
\begin{eqnarray}
{\Psi}^0   [F_2 F_2 \bar{\Phi}_{31} \bar{\phi}_{45} \phi^- h_1 + F_2 F_2 \Phi_{23} \bar{\phi}_{45} \bar{\phi}^+ h_1 
+ F_2 F_3 F_3 \phi_4 \bar{\phi}_4 \bar{\phi}_{45} \bar{f}_2 + F_4 \Phi_{23} \bar{\phi}_{45} \phi^- \bar{h}_{45} \bar{f}_5 +   
\\ \nonumber (\bar{\Phi}_{31} \bar{\phi}_{45} \phi^-  
+ \Phi_{23} \bar{\phi}_{45} \bar{\phi}^+)h_1(\bar{f}_2 l^c_2 + \bar{f}_5 l^c_5) \\
+ \nonumber \Phi_{23} \bar{\phi}_{45} \phi^- h_1 \bar{f}_1 l^c_1] 
\end{eqnarray}
\begin{eqnarray}
\bar{\Psi}^0  [F_2 F_2 \Phi_{31} \phi_{45} \bar{\phi}^- h_1 + F_2 F_2 \bar{\Phi}_{23} \phi_{45} \phi^+ h_1 +
F_2 F_2 \bar{\phi}^- h_1 h_1 \bar{h}_{45} + F_4 F_4 \Phi_{31} \phi_{45} \phi^+ h_1 + \\
\nonumber F_4 F_4 \phi^+ h_1 h_1 \bar{h}_{45} + F_4 \Phi_{31} \phi_{45} \phi^+ \bar{h}_{45} \bar{f}_5 +
+ F_4 \phi^+ h_1 \bar{h}_{45} \bar{h}_{45} \bar{f}_{5} + F_4 \bar{\phi}^- h_1 h_1 h_1 l^c_5 + \\
\nonumber (\Phi_{31} \phi_{45} \phi^+ h_1 + h_1 h_1 \bar{h}_{45})\bar{f}_1 l^c_1 + (\Phi_{31} \phi_{45} \bar{\phi}^- h_1 + \bar{\Phi}_{23} \phi_{45} \phi^+ h_1 + \bar{h}_{45} \bar{h}_{45} \bar{h}_{45} + \bar{\phi}^- h_1 h_1 \bar{h}_{45})(\bar{f}_2 l^c_2 + \bar{f}_5 l^c_5)]
\end{eqnarray}

These $10th$ order interactions would have a lifetime $\sim10^{17}-10^{52}$ years for the mass range $\sim\Lambda_4 = 10^{12}-10^{13}$~GeV and $M_s = 10^{17}-10^{18}$~GeV.  
These interactions involve multi-particle decays involving both particles and SUSY partners, within the constraints of R-parity and charge conservation.  Although there are many of these decay interactions some general comments can be made.  Almost all of them contain Higgs fields which would tend to decay (depending upon what the mass of the Higgs turns out to be) into $W^{\pm}$, quark-antiquark pairs (neutral Higgs) or $\tau$ leptons (charged Higgs), or remain as LSP if they are Higgsinos, assuming Higgsinos compose a fraction of LSP. Since the Higgs couple to heavier particles, we would expect $\bar{H}_{45}$ to decay most strongly to a pair the heaviest up-type quark allowed by kinematics, which is expected to be the c-quark. Similarly, we would expect the $H_1$ to decay most strongly to $\tau^\pm$, and to pairs of b-quarks.  Furthermore, most of the decay interactions contain many Higgs fields as well as $\mathbf{10}$ and $\mathbf{\bar{5}}$ fields which may also produce quarks and antiquarks.  Thus, several such pairs are expected to be created. These decay interactions also all involve several singlet fields which could decay into observable particles if their mass is great enough, or remain as hot-dark matter if is not.  

\section{The Fate of the Charged Tetrons}

The lifetimes and abundances of charged tetrons have recently been
discussed by Coriano {\it et.al.}~\cite{Coriano}, who have raised
questions about their lifetimes and abundances relative to those of the
neutral tetrons. In particular, they pointd out that if the only ways for
the charged tetrons to decay are through the the same higher-order
non-renormalizable operators that govern the decays of the neutral
tetrons, then, if the neutral tetrons are long-lived, so also would be the
charged tetrons, and they would probably have comparable cosmological
abundances.  Since there are very strong constraints on stable charged
matter~\cite{Perl,Ambrosio,Barish}, it was argued in~\cite{Coriano} that
tetrons could not be good candidates for dark matter.

Indeed, we do find ninth-order superpotential terms involving charged
tetrons that correspond to the annihilations of their constituents:
\begin{eqnarray}
 \bar{\Psi}^{++} && ( \Phi_{31} \bar{\phi}^-  \bar{\phi}^-  
\bar{l}^c_4  \bar{l}^c_4    
   +  \bar{\Phi}_{23}  \phi^+  \bar{\phi}^-  \bar{l}^c_4 \bar{l}^c_4), \\
\bar{\Psi}^{-}   && F_3 \phi^+   h_1   f_4   f_4  , \\
\Psi^{--} && ( \bar{\Phi}_{31} \bar{\phi}^+   \phi^-   l^c_2  l^c_2 + 
              \bar{\Phi}_{31} \bar{\phi}^+   \phi^-   l^c_5  l^c_5 
           + \bar{\Phi}_{31}  \phi^-   \phi^-   l^c_1  l^c_1 + \nonumber \\
&& \Phi_{23} \bar{\phi}^+  \bar{\phi}^+   l^c_2  l^c_2  
        \Phi_{23} \bar{\phi}^+  \bar{\phi}^+   l^c_5  l^c_5  +
        \Phi_{23} \bar{\phi}^+   \phi^-   l^c_1  l^c_1), \\
\Psi^- && ( F_1   F_1   F_3   \phi^-  \bar{h}_2 +
          F_2   F_2   F_3 \bar{\phi}^+  \bar{h}_2  + 
          F_2   F_2   F_3   \phi^-  \bar{h}_1 + \nonumber \\
&&   F_3   F_4   F_4  \phi^-  \bar{h}_2  +
	F_3   F_4   \phi^-   h_{45}  l^c_5 +
	F_3  \bar{\phi}^+ \bar{h}_2  \bar{f}_2   l^c_2 \nonumber \\
&& F_3  \bar{\phi}^+  \bar{h}_2  \bar{f}_5   l^c_5 +
	F_3   \phi^- \bar{h}_1  \bar{f}_2   l^c_2  +
	F_3 \phi^-  \bar{h}_1  \bar{f}_5   l^c_5 +
	F_3   \phi^-  \bar{h}_2  \bar{f}_1   l^c_1). \\
\bar{\Psi}^{+} && (  \bar{F}_5 \bar{F}_5 \Phi_{31} \phi^+ \bar{f}_3 +
	\Phi_{31} \bar{\phi}^- \bar{l}^c_4 \bar{l}^c_4 l^c_3 +
	 \Phi_{31} \bar{\phi}^- f_4 \bar{f}_3 \bar{l}^c_4 + 
\bar{\Phi}_{23} \phi^+ \bar{l}^c_4 \bar{l}^c_4 l^c_3).  
\end{eqnarray}
Thus, if these non-renormalizable interactions were the only ways for
charged tetrons to decay, they would have lifetimes similar to those of
the neutral tetrons. Moreover, there are no superpotential terms
corresponding to decays of $\Psi^{++}$, $\bar{\Psi}^{--}$, $\Psi^{+}$, or
$\bar{\Psi}^{-}$ states that appear before tenth order, which would
correspond to even longer lifetimes.

However, there is another mechanism which enables the heavier (charged)  
members of cryptospin multiplets to decay relatively rapidly into the
lightest (neutral) isospin partner, analogous to the $\beta$ decay of the
neutron into its lighter isospin partner in QCD, the proton.  We recall
that neutron decay is generated by a four-fermion interaction of the type
$({\bar d} u {\bar \nu} e)/m_W^2$, which leads to an effective
neutron-decay interaction of the form $({\bar n} p {\bar \nu} e)/m_W^2$.
This then leads to a neutron decay rate $\Gamma_n \sim (\delta m)^5 /
m_W^4$, where $\delta m$ is the neutron-proton mass difference. In the
case of charged-crypton decay, we expect there to exist a crypto-strong
interaction of the form
\begin{equation}
{{\bar C^+} C^0 (\pi^+ \partial \pi^0) \over \Lambda_4^2},
\label{Cplusdecay}
\end{equation}
where the $C^{+,0}$ are charged and neutral crypton fields~\footnote{In 
QCD, the $W^-$ couples to ${\bar n} p$ via a 
strongly-interacting vector meson $\rho^-$. By an analogous vector-meson 
dominance argument, one could consider the interaction (\ref{Cplusdecay}) 
as being mediated by the exchange of a `non-holomorphic' crypto-$\rho$ 
meson.}. If 
the $C^{+,0}$ mass difference $\Delta M$ were larger than $m_{\pi^+} + 
m_{\pi^0}$, the $C^+$ decay rate would be very rapid: $\Gamma_{C^+} \sim 
(\Delta M)^5 / \Lambda_4^4$. However, we expect that $\Delta M < 
m_{\pi^+}, m_{\pi^0}$, in which case the two crypto-pions must be virtual. 
In this case, the lowest-order decay interaction becomes
\begin{equation}
\alpha \Delta M {{\bar C^+} C^0 F {\tilde F} B_1 B_2 \over m_{\pi^+}^2 
m_{\pi^0}^2 \Lambda_4 M_s},
\label{Cplusvirtdecay}
\end{equation}
where $F$ denotes the Maxwell field strength and ${\tilde F}$ its dual, 
$B_{1,2}$ denote generic MSSM bosons, $M_s$ is the string 
scale, and $\alpha = \alpha(\Lambda_4)$. If the $\pi^+$ can only decay 
through higher-order interactions, 
(\ref{Cplusvirtdecay}) would be replaced by effective interactions with 
more inverse powers of $M_s$. Setting $\Delta M \sim \alpha \Lambda_4$ as 
suggested by (\ref{splittings}), and assuming the minimum values
$m_{\pi^+}^2, m_{\pi^0}^2 \sim \alpha \Lambda_4^2$ allowed by 
(\ref{pi+pi0difference}), interactions of the 
form (\ref{Cplusvirtdecay}) would yield decay rates of order
\begin{equation}
\Gamma_{C^+} \sim { \Delta M^{11} \over  \Lambda_4^8 M_s^2} \sim 
{\alpha^{11} \Lambda_4^3 \over M_s^2},
\label{Cplusrate}
\end{equation}
with additional factors of $(\Delta M / M_s)^2 \sim (\alpha \Lambda_4 /
M_s)^2$ for higher-order $\pi^+$ decay interactions. In the case of the
interactions (18) in our particular flipped SU(5) model, we would pick up
an extra factor of $(\alpha \Lambda_4 / M_s)^4$.

In this case, we estimate a charged crypton lifetime $\tau^{\pm} \sim 10^{2}- 
10^{9}$~years for $\Lambda_4 \sim 10^{13}-10^{12}$~GeV
and $M_s \sim 10^{17}$~GeV. For the same range of $\Lambda_4$ and $M_s =
10^{18}$~GeV, we estimate a charged crypton lifetime of $\tau^{\pm} \sim 
10^{8} -10^{14}$~years. These charged-tetron lifetimes are much
shorter than what we expect for the neutral tetrons. For comparison, with the same
values of $\Lambda_4$ and $M_s$, taking $M_X =
\Lambda_4$ and assuming a ninth-order
neutral-crypton decay interaction, we estimate a neutral crypton lifetime
$\tau^0 \sim 10^{13}-10^{26}$~years for $\Lambda_4 = 10^{13}-10^{12}$~GeV 
and $M_s = 10^{17}$~GeV, and $\tau^0 \sim 10^{25} - 10^{38}$
years for the same range of tetron mass and $M_s = 10^{18}$~GeV. In particular,
$\tau^{\pm} \lesssim 10^5$ years and $\tau^0 > 10^{10}$ years if
$3\cdot10^{12}~{\rm GeV} \leq \Lambda_4 \leq 2\cdot10^{13}$~GeV with $M_s = 
10^{17}$~GeV and $3\cdot10^{13}~{\rm GeV} \leq 
\Lambda_4 \leq 2\cdot10^{14}$~GeV with $M_s = 10^{18}$~GeV.
In fact, it is possible to choose a value for $
\Lambda_4$ in the expected range such that $\tau^{\pm} \lesssim 10^5$ years and
$\tau^0 > 10^{10}$ years for all values of $M_s$ between $10^{17} -
10^{18}$~GeV. Thus, it is always possible to choose reasonable values of
these parameters such that neutral tetrons will have a lifetime longer
than the present age of the universe while the charged tetrons will have
decayed prior to photon-matter decoupling. Therefore, neutral tetrons can
be in existence today as cold dark matter unencumbered by any constraints
due to charged dark matter.

\section{Conclusion}

We have made in this paper a detailed study of crypton decays in a
specific flipped SU(5) string model. We have shown that there are neutral
tetrons that are naturally metastable in this string model, with lifetimes
long enough to make perfect candidates for cold dark matter and possibly
act as sources of UHECRs. Moreover, their charged `cryptospin' partners
naturally decay much more rapidly, with lifetimes that may be much shorter
than the age of the Universe. Thus, the flipped SU(5) string model does
not predict the existence of any charged cold dark matter. Time will tell
whether the UHECRs are in fact due to the decays of ultraheavy particles,
but the flipped SU(5) string model seems to have the appropriate
characteristics for this to be possible, as well as providing possible
cold dark matter candidates in the forms of its neutral tetron bound
states.  We believe that these properties along with the other successes
of string-derived flipped SU(5) such as dynamic double-triplet splitting and natural 
suppression of dimension-5 operators that mediate rapid proton decay make this 
model particularly attractive and should strongly motivate future study.  
 
\section*{Acknowledgements}

The work of D.V.N. is supported by D.O.E. grant DE-FG03-95-ER-40917. V.E.M 
would like to thank G. Cleaver and J. Walker for their early help and valuable discussions.

\end{document}